\documentclass[aps,showpacs,onecolumn,10pt]{revtex4}
\usepackage{amsfonts}
\usepackage{amsmath}
\usepackage{amssymb}
\usepackage{graphicx}
\usepackage[latin1]{inputenc}
\usepackage[T1]{fontenc}

\begin{document}

\title{A linearized kinetic theory of spin-1/2 particles in magnetized
plasmas}
\author{J. Lundin and G. Brodin}
\affiliation{Department of Physics, Ume\aa University, SE--901 87 Ume\aa, Sweden}

\begin{abstract}
We have considered linear kinetic theory including the electron spin
properties in a magnetized plasma. The starting point is a mean field
Vlasov-like equation, derived from a fully quantum mechanical treatment,
where effects from the electron spin precession and the magnetic dipole
force is taken into account. The general conductivity tensor is derived,
including both the free current contribution, as well as the magnetization
current associated with the spin contribution. We conclude the paper with an
extensive discussion of the quantum-mechanical boundary where we list
parameter conditions that must be satisfied for various quantum effects to
be influential.
\end{abstract}

\maketitle
\section{Introduction}

Recently there has been much interest in the properties of quantum plasmas,
see e.g.\ Refs.\ \cite%
{Manfredi-review,Shukla-Eliasson-review,Haas-2000,Garcia-2005,Shukla-Eliasson-2006,Jens2009,Marklund2007,Classical-quant,Marklund-plasmonic,Brodin2008,BMZ-2009,Manfredi-quantum-well}%
. The research has been motivated by applications to e.g.\ quantum wells 
\cite{Manfredi-quantum-well}, plasmonics \cite%
{Atwater-Plasmonics,Marklund-plasmonic}, spintronics \cite{Spintronics},
astrophysics \cite{Astro} and ultra-cold plasmas \cite{Ultracold}. Quantum
plasma effects has also been measured in solid density target experiments 
\cite{Solid-density}. Within fluid theory, the theoretical models applied
cover effects such as particle dispersion (the Bohm-de Broglie potential),
and Fermi pressure. The magnetic dipole force and the magnetization current
associated with the electron spin has also been captured within fluid
models, with a macroscopic spin density as an extra dependent variable \cite%
{Marklund2007}. More accurate models applying kinetic theory has also been
studied \cite{Cowley-1986,Kulsrud-1986,Brodin2008,Jens2009}. In the absence
of spin effects, a kinetic evolution equation can be derived for the Wigner
function \cite{Manfredi-review}. For long scale-lengths this equation
reduces to the classical Vlasov equation.

Including the spin degrees of freedom, it was recently shown \cite%
{Jens2009} that a Wigner transform in regular phase space and a Q-transform
in spin space produces a physically attractive evolution equation for a
scalar distribution function. If the magnetic moment of the particles is put
to zero, the ordinary Wigner equation is immediately recovered. Here we will
study the opposite limit, however. For wavelengths much longer than the
characteristic de Broglie length of the particles, the quantum effects
associated with particle dispersive effects (de-localized wavefunctions)
disappear, and the evolution equation is much simplified. The remaining
quantum effects are all due to the electron spin, explicitly due to terms
being proportional to the magnetic moment in the dynamic equations, or
indirectly due to the background distribution function obeying Fermi-Dirac
statistic. The kinetic model obtained from the long-scale limit of a fully
quantum mechanical treatment is a slight generalization of a semi-classical
spin model presented in Ref.\ \cite{Brodin2008}, where the spin vector
complements the regular phase-space variables as an independent variable 
\cite{Brodin2008,Jens2009}. In Ref.\ \cite{Brodin2008} it was also shown
that new wave modes appear in a magnetized plasma due to the combined
dynamics of the magnetic dipole force and the spin precession. Here we will
extend the linear analysis made in that work in several directions:

\begin{enumerate}
\item By considering the case of general wave propagation (i.e.\ an
arbitrary direction of propagation and a general wave polarization).

\item By not making any restrictions on the wave frequency domain. 

\item By using a somewhat more accurate model, derived from a fully quantum
mechanical treatment.

\item By presenting a more thorough discussion of the classical - quantum
mechanical boundary.
\end{enumerate}

The theory turns out to produce a Hermitian structure for the general
conductivity tensor, including the free and magnetization currents, where
all non-Hermitian contribution are associated with the poles. Contrary to
common quantum effects associated with particle dispersion (de-localized
wave functions), the effects due to the electron spin does not necessarily
vanish in a plasma of moderate density and temperature. A low temperature
and a high density tend to make spin effects more important (similarly to
the quantum effects described in e.g.\ Refs.\ \cite%
{Manfredi-review,Shukla-Eliasson-review}), but\ a strong magnetic field or
wave particle resonances can modify this picture considerably. As will be
discussed in some detail in Section III, we point out that there is a large
number of dimensionless parameters that together captures the relative
significance of spin effects in different plasma regimes. Furthermore, in
order to demonstrate the usefulness of the general conductivity tensor, in
Section IV we present an analysis of the transition from wave damping to
instability, when free energy is added to the the spin-degrees of freedom in
the background distribution function. Specifically, it is shown that a
background distribution function with equal population of the spin-up and
spin-down states states leads to an instability. 

\section{The conductivity tensor}

For long spatial scale-length the evolution of a (quasi) distribution
function $f(\mathbf{r,v,}\hat{\mathbf{s}},t)$ for electrons is described by
the Vlasov-like equation \cite{Jens2009} 
\begin{equation}
\frac{\partial f}{\partial t}+\mathbf{v}\cdot \nabla _{\mathbf{x}}f+\left[ %
\frac{q_e}{m_e}(\mathbf{E}+\mathbf{v}\times \mathbf{B})+\frac{\mu _{e}}{m_e}%
\nabla _{\mathbf{x}}\left( \hat{\mathbf{s}}\cdot \mathbf{B}+\mathbf{B}\cdot
\nabla _{\hat{\mathbf{s}}}\right) \right] \cdot \nabla _{\mathbf{v}}f+\frac{%
2\mu _{e}}{\hbar }(\hat{\mathbf{s}}\times \mathbf{B})\cdot \nabla _{\hat{%
\mathbf{s}}}f=0,  \label{Vlasov}
\end{equation}%
where $q_e = -e$ is the charge of an electron, $\mu _{e}=-(g/2)e\hbar /2m_e$ is the electron magnetic moment (note
that the sign is included) and $g=2.002319$ is the electron spin $g$ factor, 
$\hat{\mathbf{s}}$ is the unit spin vector, $\hbar $ is Planck's constant
divided by $2\pi $, and $m_e$ is the electron mass. The distribution function $f$ is normalized such that $\int
fd^{2}sd^{3}v=n$ with $n$ being the number density. As for spin variables it
is convenient to use ordinary spherical coordinates, i.e.\ $d^{2}s=\sin
\theta _{s}d\theta _{s}d\varphi _{s}$.

In the derivation of Eq.\ (\ref{Vlasov}) found in Ref.\ \cite{Jens2009},
trigonometric operators are expanded in powers of $\hbar $. Consequently,
Eq.\ (\ref{Vlasov}) is the long scale limit of a more complete quantum
description, where long scale here means that length scales are larger than
the characteristic de Broglie wavelength of the electrons. Upon disregarding
spin effects, the opposite limit of the full quantum theory, i.e.\ for short
length scales, is the well-known evolution equation for the Wigner function 
\cite{Manfredi-review}. The spin term proportional to $\hat{\mathbf{s}}\cdot 
\mathbf{B}$ in Eq.\ (\ref{Vlasov}) is due to the magnetic dipole force, and
the term proportional to $\hat{\mathbf{s}}\times \mathbf{B}$ owes to the
spin precession. Both these terms are also found in the semi-classical
version of Eq.\ (\ref{Vlasov}), see Ref.\ \cite{Brodin2008}. The spin term
proportional to $\mathbf{B}\cdot \nabla _{\hat{\mathbf{s}}}$, however, is a
fully quantum mechanical effect. Essentially this term can be viewed as a
modification of the magnetic dipole force that occurs due to the spread out
nature of the spin probability distribution.

By solving Eq.\ (\ref{Vlasov}) we may find an expression for the
distribution function $f$ from which we can construct the free current, $%
\mathbf{J}_f$, and the magnetization current density, $\mathbf{J}_M$,
according to \cite{Jens2009} 
\begin{eqnarray}
\mathbf{J} &=&\mathbf{J}_{f}+\mathbf{J}_{M}  \notag  \label{J} \\
&=&\mathbf{J}_{f}+\nabla _{\mathbf{x}}\times \mathbf{M}  \notag \\
&=&q_e\int \mathbf{v}f d^{2}sd^{3}v+\nabla _{\mathbf{x}}\times \left( \mu
_{e}\int 3\hat{\mathbf{s}}f d^{2}sd^{3}v\right) .
\end{eqnarray}%
In this section we only consider the electron contribution to the current
density. The ion contribution may typically be treated classically (because
ion spin effects are suppressed compared to electron spin effects due the
heavy ion mass), and its contribution may be found in standard literature,
see e.g.\ Ref.\ \cite{Swanson}. The derivation given below may, however,
easily be generalized to also apply for arbitrary spin-1/2 particles through
the appropriate substitution of charge, mass and magnetic moment.

Equations (\ref{Vlasov}) and (\ref{J}) gives, together with the Maxwell's
equations, a closed system of equations that can be solved to describe the
dynamics of a plasma where quantum effects of the electron spin are
captured. Below we will derive the conductivity tensor $\sigma _{ij}$,
defined as $J_{i}=\sigma _{ij}E_{j}$, for such a system \cite{note1}. With
the conductivity tensor known, it is straightforward to construct a
dispersion matrix and find the dispersion relations for arbitrary wave
modes. Next, we linearize the kinetic Eq.\ (\ref{Vlasov}) according to $%
f=f_{0}+f_{1}$ and $\mathbf{B}=\mathbf{B}_{0}+\mathbf{B}_{1}$, where the
subscript $0$ denotes an unperturbed quantity and the subscript $1$ denotes
a perturbation, and we take $\mathbf{B}_{0}=B_{0}\hat{\mathbf{z}}$. Before
proceeding, let us point out a few quantum effects that may be contained
already in the unperturbed distribution function:

\begin{enumerate}
\item \textit{Fermi-Dirac statistics:} This effect is well-known. Here we
just note that for a plasma of moderate density and temperature this effect
may be suppressed, as a large (negative) chemical potential (that applies
for $\hbar ^{2}n_{0}^{2/3}/m_{e}k_{B}T\ll 1$) turns the thermodynamic
equilibrium distribution into the classical Maxwellian.

\item \textit{Landau-quantization:} The quantization of perpendicular energy
states becomes important in the regime of very strong magnetic fields, or
very low temperatures, when $\hbar \left\vert \omega _{ce}\right\vert
/k_{B}T\rightarrow 1$, where $k_{B}$ is Boltzmann's constant and $\omega
_{ce}=-eB_{0}/m_e$ is the electron cyclotron frequency.

\item \textit{Spin-splitting: }The two spin states, up- and down relative to
the magnetic field, clearly have different probability distributions in spin
space. As a result, the general time-independent distribution function can
be written as $f_{0}=f_{0+}+f_{0-}$ with $f_{0\pm }=(1/4\pi )F_{0\pm }(%
\mathbf{v})(1\pm \cos \theta _{s})$, where for a time-independent
distribution function $F_{0\pm }$ can be arbitrary functions of $(v_{\perp
},v_{z})$, and $F_{0\pm }$ is normalized such that $\int F_{0\pm
}d^{3}v=n_{0\pm }$ with $n_{0\pm }$ being the number densities of the spin
up/down states respectively. The positive spin state here means that the
spin points in the direction parallel to the magnetic field, which means
that the magnetic moment points in the opposite direction. Note that with
this definition, the lower energy state is the spin state with negative
index, i.e. $n_{0-}>n_{0+}$ in case the background distribution $f_{0}$
describes a thermodynamic equilibrium.
\end{enumerate}

For a full quantum mechanical expression of the thermodynamic equilibrium
distribution, see Ref.\ \cite{Jens2009}. However, in the present manuscript
we will focus on the effects due to the dynamical equations, rather than the
initial conditions briefly discussed above. Still, we note that since the
spin-distribution has no classical limit point 3 above cannot be ignored.
Since the present collision-free model has no mechanism for spin-flips to
occur, we also note that the number of particles with spin-up $N_{+}=\int
f_{0+}d^{2}sd^{3}vd^{3}r$ and the number of particles with spin down $%
N_{-}=\int f_{0-}d^{2}sd^{3}vd^{3}r$ are conserved quantities. Furthermore,
in thermodynamic equilibrium the spin-up and spin-down number densities $%
n_{0\pm }=\int f_{0\pm}d^{2}sd^{3}v$ are related by $(n_{0+} -
n_{0-})/(n_{0+} + n_{0-})=\tanh (\mu_{e}B_{0}/k_{B}T)$. Since the evolution
equation is the same, independent of the unperturbed spin state, it is
convenient to wait with the split of $f_{0}$ into $f_{0\pm }$ to the final
stages of the calculation.

After linearization, Eq.\ (\ref{Vlasov}) is written as 
\begin{eqnarray}
&& \frac{\partial f_{1}}{\partial t}+\mathbf{v}\cdot \nabla _{\mathbf{x}%
}f_{1}+\frac{q_e}{m_e}(\mathbf{v}\times \mathbf{B}_{0})\cdot \nabla _{\mathbf{v%
}}f_{1}+\frac{2\mu _{e}}{\hbar }(\hat{\mathbf{s}}\times \mathbf{B}_{0})\cdot
\nabla _{\hat{\mathbf{s}}}f_{1}  \notag  \label{linVlasov} \\
&=&-\left[ \frac{q_e}{m_e}(\mathbf{E}+\mathbf{v}\times \mathbf{B}_{1})+\frac{%
\mu _{e}}{m_e}\nabla _{\mathbf{x}}\left( \hat{\mathbf{s}}\cdot \mathbf{B}%
_{1}+\mathbf{B}_{1}\cdot \nabla _{\hat{\mathbf{s}}}\right) \right] \cdot
\nabla _{\mathbf{v}}f_{0}-\frac{2\mu _{e}}{\hbar }(\hat{\mathbf{s}}\times 
\mathbf{B}_{1})\cdot \nabla _{\hat{\mathbf{s}}}f_{0}.
\end{eqnarray}%
Next we make a plane wave ansatz of the perturbed parameters according to $%
f_{1}=\tilde{f}_{1}\exp [i(\mathbf{k}\cdot \mathbf{x}-\omega t)]$, etc.
Without loss of generality, we define the wavevector as $\mathbf{k} = k_\bot%
\hat{\mathbf{x}} + k_z \hat{\mathbf{z}}$. We also choose to express the
velocity in cylindrical coordinates $(v_{\perp },\varphi _{v},v_{z})$ such
that $d^{3}v=v_{\bot }dv_{\bot }d\varphi _{v}dv_{z}$, and expand $f_{1}$ in
eigenfunctions to the operator of the right hand side 
\begin{equation}
\tilde{f}_{1}=\sum_{a=-\infty}^{\infty}\sum_{b=-\infty}^{\infty}g_{ab}(v_{%
\perp },v_{z},\theta _{s})\psi _{a}(\varphi _{v},v_{\perp })\frac{1}{\sqrt{%
2\pi }}\exp (-ib\varphi _{s}),  \label{f1_expansion}
\end{equation}%
where 
\begin{eqnarray}
\psi _{a}(\varphi _{v},v_{\perp }) &=&\frac{1}{\sqrt{2\pi }}\exp
[-i(a\varphi _{v}-k_{\perp }v_{\perp }\sin \varphi _{v}/\omega_{ce})]  \notag
\label{psi} \\
&=&\frac{1}{\sqrt{2\pi }}\sum_{l=-\infty}^\infty\mathcal{J}_{l}\left( \frac{%
k_{\perp }v_{\perp }}{\omega_{ce}}\right) \exp [i(l-a)\varphi _{v}],
\end{eqnarray}%
where $\mathcal{J}_l(x)$ is a Bessel function of the first kind. We may then
note the following simplifying relations; 
\begin{eqnarray}
\frac{q_e}{m_e}(\mathbf{v}\times \mathbf{B}_{0})\cdot \nabla _{\mathbf{v}%
}f_{1} &=&-\omega _{ce}\frac{\partial f_{1}}{\partial \varphi _{v}},  \notag
\\
\frac{2\mu _{e}}{\hbar }(\hat{\mathbf{s}}\times \mathbf{B}_{0})\cdot \nabla
_{\hat{\mathbf{s}}}f_{1} &=&-\omega _{cg}\frac{\partial f_{1}}{\partial
\varphi _{s}},
\end{eqnarray}%
where $\omega_{cg}=2\mu_eB_0/\hbar$ is the spin precession frequency.
Moreover, for simplicity we assume an isotropic distribution function on the
form $f_{0}=f_{0}(v^{2},\theta _{s})$. For this case we may drop the term $(%
\mathbf{v}\times \mathbf{B}_{1})\cdot \nabla _{\mathbf{v}}f_{0}$. We may
also take advantage of the relations $(\partial f_{0}/\partial v_{\bot
})/2v_{\bot }=(\partial f_{0}/\partial v_{z})/2v_{z}=\partial f_{0}/\partial
v^{2}$.

Using the eigenfunction expansion of $\tilde{f}_{1}$ (Eq.\ (\ref%
{f1_expansion})) in the linearized Vlasov equation (\ref{linVlasov}),
multiplying the resulting equation with $\psi _{a}^{\ast }e^{ib\varphi _{s}}/%
\sqrt{2\pi }$ (where the star denotes complex conjugate) and integrating over 
$\varphi _{v}$ and $\varphi _{s}$, we find the equation 
\begin{subequations}
\begin{equation}
i(\omega -k_{z}v_{z}-a\omega _{ce}-b\omega _{cg})g_{ab}=I_{ab}(v_{\perp
},v_{z},\varphi _{s}),  \label{gab}
\end{equation}%
where 
\begin{equation}
I_{ab}=\int_{0}^{2\pi }\int_{0}^{2\pi }d\varphi _{v}d\varphi _{s}\left[ \left( \frac{q_e}{%
m_{e}}\tilde{\mathbf{E}}+\frac{\mu _{e}}{m_{e}}\nabla _{\mathbf{x}}\left( 
\hat{\mathbf{s}}\cdot \tilde{\mathbf{B}}_{1}+\tilde{\mathbf{B}}_{1}\cdot
\nabla _{\hat{\mathbf{s}}}\right) \right) \cdot \nabla _{\mathbf{v}}f_{0}+%
\frac{2\mu _{e}}{\hbar }(\hat{\mathbf{s}}\times \tilde{\mathbf{B}}_{1})\cdot
\nabla _{\hat{\mathbf{s}}}f_{0}\right] \psi _{a}^{\ast }\frac{1}{\sqrt{2\pi }%
}\exp (ib\varphi _{s}).  \label{Iab}
\end{equation}%
From Eq.\ (\ref{gab}) we can then solve $g_{ab}$ and thereby construct $%
\tilde{f}_{1}$ in terms of $f_{0}$. It may be useful to note that the vector
products in (\ref{Iab}) are; 
\end{subequations}
\begin{subequations}
\begin{eqnarray}
\tilde{\mathbf{E}}\cdot \nabla _{\mathbf{v}}f_{0} &=&2\left( E_{x}v_{\bot
}\cos \varphi _{v}+E_{y}v_{\bot }\sin \varphi _{v}+E_{z}v_{z}\right) \frac{%
\partial f_{0}}{\partial v^{2}},  \label{h1} \\
\nabla _{\mathbf{x}}\left( \hat{\mathbf{s}}\cdot \tilde{\mathbf{B}}%
_{1}\right) \cdot \nabla _{\mathbf{v}}f_{0} &=&2i\left( k_{z}v_{z}+k_{\bot
}v_{\bot }\cos \varphi _{v}\right) \left( B_{x}\sin \theta _{s}\cos \varphi
_{s}+B_{y}\sin \theta _{s}\sin \varphi _{s}+B_{z}\cos \theta _{s}\right) 
\frac{\partial f_{0}}{\partial v^{2}},  \label{h2} \\
\nabla _{\mathbf{x}}\left( \tilde{\mathbf{B}}_{1}\cdot \nabla _{\hat{\mathbf{%
s}}}\right) \cdot \nabla _{\mathbf{v}}f_{0} &=&2i\left( k_{z}v_{z}+k_{\bot
}v_{\bot }\cos \varphi _{v}\right) \left( B_{x}\cos \theta _{s}\cos \varphi
_{s}+B_{y}\cos \theta _{s}\sin \varphi _{s}-B_{z}\sin \theta _{s}\right) 
\frac{\partial ^{2}f_{0}}{\partial \theta _{s}\partial v^{2}},  \label{h3} \\
(\hat{\mathbf{s}}\times \tilde{\mathbf{B}}_{1})\cdot \nabla _{\hat{\mathbf{s}%
}}f_{0} &=&\left( B_{x}\sin \varphi _{s}-B_{y}\cos \varphi _{s}\right) \frac{%
\partial f_{0}}{\partial \theta _{s}}.  \label{h4}
\end{eqnarray}%
With the relations (\ref{psi}) and (\ref{h1})-(\ref{h4}), the integration in
(\ref{Iab}) can be performed in a tedious but straight forward manner (some
useful integrals are found in appendix \ref{appendix:relations}, Eqs.\ (\ref%
{inthelp1})-(\ref{inthelp3})), and through Eq.\ (\ref{f1_expansion}) the
resulting expression for $\tilde{f}_{1}$ becomes 
\end{subequations}
\begin{equation}
\tilde{f}_{1}=\sum_{a=-\infty }^{\infty }\sum_{l=-\infty }^{\infty }\mathcal{%
J}_{l}e^{i(l-a)\varphi _{v}}\left( \frac{Ae^{i\varphi _{s}}}{\omega
-k_{z}v_{z}-a\omega _{ce}+\omega _{cg}}+\frac{B}{\omega -k_{z}v_{z}-a\omega
_{ce}}+\frac{Ce^{-i\varphi _{s}}}{\omega -k_{z}v_{z}-a\omega _{ce}-\omega
_{cg}}\right)   \label{f1}
\end{equation}%
with 
\begin{eqnarray}
A &=&\frac{\mu _{e}}{m_{e}\omega }\left( -ik_{z}E_{x}-k_{z}E_{y}+ik_{\bot
}E_{z}\right) \left[ -\frac{m_{e}}{\hbar }\frac{\partial f_{0}}{\partial
\theta _{s}}+\left( a\omega _{ce}+k_{z}v_{z}\right) \left( \sin \theta _{s}%
\frac{\partial f_{0}}{\partial v^{2}}+\cos \theta _{s}\frac{\partial
^{2}f_{0}}{\partial \theta _{s}\partial v^{2}}\right) \right] \mathcal{J}%
_{a},  \notag \\
B &=&2\frac{\mu _{e}}{m_{e}\omega }E_{y}k_{\bot }\left( a\omega
_{ce}+k_{z}v_{z}\right) \left( \cos \theta _{s}\frac{\partial f_{0}}{%
\partial v^{2}}-\sin \theta _{s}\frac{\partial ^{2}f_{0}}{\partial \theta
_{s}\partial v^{2}}\right) \mathcal{J}_{a}-2i\frac{q_e}{m_{e}}\left( E_{x}a%
\frac{\omega _{ce}}{k_{\bot }}\mathcal{J}_{a}+iE_{y}v_{\bot }\mathcal{J}%
_{a}^{\prime }+E_{z}v_{z}\mathcal{J}_{a}\right) \frac{\partial f_{0}}{%
\partial v^{2}},  \notag \\
C &=&\frac{\mu _{e}}{m_{e}\omega }\left( ik_{z}E_{x}-k_{z}E_{y}-ik_{\bot
}E_{z}\right) \left[ \frac{m_{e}}{\hbar }\frac{\partial f_{0}}{\partial
\theta _{s}}+\left( a\omega _{ce}+k_{z}v_{z}\right) \left( \sin \theta _{s}%
\frac{\partial f_{0}}{\partial v^{2}}+\cos \theta _{s}\frac{\partial
^{2}f_{0}}{\partial \theta _{s}\partial v^{2}}\right) \right] \mathcal{J}%
_{a}.  \notag
\end{eqnarray}%
Here, the argument of the Bessel functions is $k_{\bot }v_{\bot }/\omega
_{ce}$. In obtaining Eq.\ (\ref{f1}) we have used $\nabla \times \mathbf{E}%
=-\partial \mathbf{B/}\partial t$ to relate the magnetic field to the
electric field, as well as the Bessel function identities $\mathcal{J}%
_{a+1}(x)+\mathcal{J}_{a-1}(x)=2a\mathcal{J}_{a}(x)/x$ and $\mathcal{J}%
_{a-1}(x)-\mathcal{J}_{a+1}(x)=2\mathcal{J}_{a}^{\prime }(x)$.

With $\tilde{f}_{1}$ known in terms of $f_{0}$, we proceed by splitting $%
f_{0}$ into its spin states, $f_{0}=f_{0+}+f_{0}$, and construct the free
current 
\begin{eqnarray}
\mathbf{J}_{f} &=&\sum_{\nu =+,-}q_e\int \mathbf{v}f_{1\nu }d^{2}sd^{3}v 
\notag  \label{Jf} \\
&=&\sum_{\nu =+,-}q_e\int \sum_{a=-\infty }^{\infty }\left[ a\frac{\omega
_{ce}}{k_{\bot }}\mathcal{J}_{a}\hat{\mathbf{x}}-iv_{\bot }\mathcal{J}%
_{a}^{\prime }\hat{\mathbf{y}}+v_{z}\mathcal{J}_{a}\hat{\mathbf{z}}\right] 
\tilde{B}_{\nu }d^{2}sd^{3}v
\end{eqnarray}%
as well as the magnetization current 
\begin{eqnarray}
\mathbf{J}_{M} &=&\sum_{\nu =+,-}\nabla \times \left( 3\mu _{e}\int \mathbf{s%
}f_{1\nu }d^{2}sd^{3}v\right)   \notag  \label{Jm} \\
&=&\sum_{\nu =+,-}\frac{3}{2}\mu _{e}\int \sum_{a=-\infty }^{\infty }%
\mathcal{J}_{a}\sin \theta _{s}\left[ \left( \tilde{A}_{\nu }-\tilde{C}_{\nu
}\right) k_{z}\hat{\mathbf{x}}+i\left( \left( \tilde{A}_{\nu }+\tilde{C}%
_{\nu }\right) k_{z}-\frac{2\cos \theta _{s}}{\sin \theta _{s}}\tilde{B}%
_{\nu }k_{\bot }\right) \hat{\mathbf{y}}-\left( \tilde{A}_{\nu }-\tilde{C}%
_{\nu }\right) k_{\bot }\hat{\mathbf{z}}\right] d^{2}sd^{3}v,  \notag \\
&&
\end{eqnarray}%
where we have performed the $\varphi _{v}$ and $\varphi _{s}$ integration
and immediately reintroduced the full integration element for convenience of
a compact notation. We have also included the denominators in (\ref{f1}) in $%
\tilde{A}_{\nu }$ by defining $\tilde{A}_{\nu }=A_{\nu
}/(\omega -k_{z}v_{z}-a\omega _{ce}+\omega _{cg})$, and similarly for $%
\tilde{B}_{\nu }$ and $C_{\nu }$.

Before we construct the conductivity tensor, we note that we can simplify
things yet further by writing $f_{0\pm }=(1/4\pi )(1\pm \cos \theta
_{s})F_{0\pm }(v^2)$. We may then perform the $\theta _{s}$ integrations in
Eqs.\ (\ref{Jf}) and (\ref{Jm}) (a useful integral is found in Appendix \ref%
{appendix:relations}, Eq.\ (\ref{inthelp4})), and construct the conductivity
tensor which may be written as 
\begin{equation}
\sigma _{ij}=\sum_{\nu =+,-}\sum_{a=-\infty }^{\infty }\int \left[ \frac{%
X_{(\nu)ij}^{(\text{sp})}}{\omega -k_{z}v_{z}-a\omega _{ce}+\omega _{cg}}+%
\frac{Y_{(\nu)ij}^{(\text{cl})}+Y_{(\nu)ij}^{(\text{sp})}}{\omega
-k_{z}v_{z}-a\omega _{ce}}+\frac{Z_{(\nu)ij}^{(\text{sp})}}{\omega
-k_{z}v_{z}-a\omega _{ce}-\omega _{cg}}\right] d^{3}v  \label{sigma}
\end{equation}%
where 
\begin{equation}
Y_{(\nu)ij}^{(\text{cl})}=2\frac{q_e^{2}}{m_e}\frac{\partial F_{0\nu }}{%
\partial v^{2}}\times \left( 
\begin{array}{ccc}
-ia^{2}\frac{\omega _{ce}^{2}}{k_{\bot }^{2}}\mathcal{J}_{a}^{2} & a\frac{%
\omega _{ce}}{k_{\bot }}v_{\bot }\mathcal{J}_{a}\mathcal{J}_{a}^{\prime } & 
-ia\frac{\omega _{ce}}{k_{\bot }}v_{z}\mathcal{J}_{a}^{2} \\ 
&  &  \\ 
-a\frac{\omega _{ce}}{k_{\bot }}v_{\bot }\mathcal{J}_{a}\mathcal{J}%
_{a}^{\prime } & -iv_{\bot }^{2}\mathcal{J}_{a}^{^{\prime }2} & -v_{\bot
}v_{z}\mathcal{J}_{a}\mathcal{J}_{a}^{\prime } \\ 
&  &  \\ 
-ia\frac{\omega _{ce}}{k_{\bot }}v_{z}\mathcal{J}_{a}^{2} & v_{\bot }v_{z}%
\mathcal{J}_{a}\mathcal{J}_{a}^{\prime } & -iv_{z}^{2}\mathcal{J}_{a}^{2}%
\end{array}%
\right)  \notag
\end{equation}%
is the classical contribution, and the spin contributions are 
\begin{equation}
Y_{(\nu)ij}^{(\text{sp})}=-2\mu _{e}\frac{q_e}{m_e}\frac{\partial F_{0\nu }}{%
\partial v^{2}}\times \left( 
\begin{array}{ccc}
0 & -\nu a\omega _{ce}\mathcal{J}_{a}^{2} & 0 \\ 
&  &  \\ 
\nu a\omega _{ce}\mathcal{J}_{a}^{2} & i\frac{\mu _{e}}{q_e}\frac{k_{\bot
}^{2}}{\omega }\left( a\omega _{ce}+k_{z}v_{z}\right) \mathcal{J}_{a}^{2} & 
\nu k_{\bot }v_{z}\mathcal{J}_{a}^{2} \\ 
& +\nu i\left( \omega +a\omega _{ce}+k_{z}v_{z}\right) \frac{k_{\bot
}v_{\bot }}{\omega }\mathcal{J}_{a}\mathcal{J}_{a}^{\prime } &  \\ 
&  &  \\ 
0 & -\nu k_{\bot }v_{z}\mathcal{J}_{a}^{2} & 0%
\end{array}%
\right)  \notag
\end{equation}%
together with 
\begin{eqnarray}
X_{(\nu)ij}^{(\text{sp})} &=&i\frac{\mu _{e}^{2}}{\hbar \omega }\left( \nu
F_{0\nu }+\frac{\hbar }{m_e}\left( a\omega _{ce}+k_{z}v_{z}\right) \frac{%
\partial F_{0\nu }}{\partial v^{2}}\right) \mathcal{J}_{a}^{2}M_{ij}  \notag
\\
Z_{(\nu)ij}^{(\text{sp})} &=&i\frac{\mu _{e}^{2}}{\hbar \omega }\left( -\nu
F_{0\nu }+\frac{\hbar }{m_e}\left( a\omega _{ce}+k_{z}v_{z}\right) \frac{%
\partial F_{0\nu }}{\partial v^{2}}\right) \mathcal{J}_{a}^{2}M^*_{ij} 
\notag
\end{eqnarray}%
where 
\begin{equation}
M_{ij}=\left( 
\begin{array}{ccc}
-k_{z}^{2} & ik_{z}^{2} & k_{\bot }k_{z} \\ 
&  &  \\ 
-ik_{z}^{2} & -k_{z}^{2} & ik_{\bot }k_{z} \\ 
&  &  \\ 
k_{\bot }k_{z} & -ik_{\bot }k_{z} & -k_{\bot }^{2}%
\end{array}%
\right) .  \notag
\end{equation}%
Given the conductivity tensor (\ref{sigma}), the general dispersion relation
is obtained in the same way as in the classical case, i.e.\ $\text{det}%
D_{ij}=0$, with $D_{ij} =
\delta_{ij}-k_ik_jc^2/\omega^2-i\sigma_{ij}/\varepsilon_0\omega$.
Analogously to the classical case the conductivity tensor has a Hermitian
structure ($\sigma _{ij}=-\sigma_{ji}^{\ast }$) if the pole-contributions
(associated with the denominators $(\omega-k_zv_z-a\omega_{ce}-b\omega_{cg})$%
) are dropped. It should be noted the Hermitian structure of the
conductivity tensor does not follow trivially from Eqs.\ (\ref{Jf}) and (\ref%
{Jm}). A detailed discussion on this is found in appendix \ref{App-Hermitian}%
.

In appendix \ref{sec:SLRL-limit}, we derive the short Larmor radius limit of
the conductivity tensor (\ref{sigma}), i.e.\ the limit in which the argument
of the Bessel functions are small, $k_{\bot }v_{\bot }/\omega _{ce}\ll 1$.


\section{The classical - quantum mechanical boundary}

In addition to the spin phenomena studied here, using Eq.\ (\ref{Vlasov}),
kinetic quantum plasma phenomena associated with non-localized wave
functions can be studied using the Wigner function, see e.g.\ Ref.\ \cite%
{Manfredi-review}. Those quantum effects prove to be important for
wavelengths shorter than the thermal de Broglie wavelength. However,
collective effects are typically negligible for such short wavelengths,
unless the parameter $d_{1}=\hbar \omega _{p}/k_{B}T_{\mathrm{m}}$
approaches unity, where $\omega_p = (n_0e^2/\varepsilon_0m_e)^{1/2}$ is the
plasma frequency. Here $T_{\mathrm{m}}$ is the largest quantity of the Fermi
temperature $T_{\mathrm{F}}$ and the thermodynamic temperature $T.$ Thus it
is safe to say that the quantum effects that are \textit{not} associated
with the spin requires either a low temperature and/or a high density. In
addition to quantum effects associated with de-localized wavefunctions,
there are those associated with the Fermi pressure. Naturally the relative
importance of this effect is captured by the parameter $d_{2}=T_{\mathrm{F}%
}/T=(3\pi ^{2})^{2/3}\hbar ^{2}n^{2/3}/2m_ek_{B}T$, which also requires a
low temperature and/or a high density plasma, although the scaling differs
somewhat from that of $d_{1}$. In linear theory, given a classical kinetic
dispersion relation for a thermodynamic equilibrium system, the effects of
the Fermi pressure is relatively trivial, since it can be deduced simply by
replacing an unperturbed Maxwellian distribution with the Fermi-Dirac
distribution. Estimating when the electron spin effects is of importance is
somewhat more complicated, as the relevant dimensionless parameter now also
involves the magnetic field strength and the wave-number and wave-frequency.
As will be verified below, from the expression for the conductivity tensor
components (\ref{sigma}), we point out a set of important parameters
determining the relative strength of spin effects. Throughout this section
we will assume the presence of ions described by a classical contribution to
the conductivity tensor, see e.g.\ Ref.\ \cite{Swanson} and electrons
described by Eq.\ (\ref{sigma}).

\begin{itemize}
\item The Zeeman energy over thermal energy, $d_{3}=\mu
_{e}B_{0}/k_{B}T=\hbar \omega _{ce}/k_{B}T$. This parameters enters for
several reasons. The simplest one is that it determines the ratio between
the spin-up and -down populations (i.e.\ $F_{0+}$ versus $F_{0-}$ in (\ref%
{sigma})) in thermodynamic equilibrium.

\item $d_{4}=\hbar k^{2}/m_{e}\omega $. This parameter is important since it
determines the relative strength of the magnetic dipole force as compared to
the force from the electric field. For systems that allow short-wavelength
low-frequency modes this is of particular significance.

\item $d_{5}=\hbar \omega _{ce}/m_{i}c_{A}^{2}$, where $c_{A}=(B_{0}^{2}/\mu
_{0}m_{i}n_{0})^{1/2}$ is the Alfv\'en velocity. This parameter can appear
for several reasons. A relatively simple example is that it gives the
relative strength of the magnetic dipole force to the Lorentz force,
evaluated for ion-cyclotron waves.

\item $d_{6}=$ $\hbar \omega _{p}/m_{e}c^{2}$. \ This parameter describes
the relative magnetic permeability due to the spin, for a given spin state.
\ When it approaches unity the density is so high that a relativistic
treatment is called for. Thus the regime $d_{6}>1$ is strictly not allowed
by our treatment. However, as we will see below, the parameter $d_{6}$ still
have a certain significance, also for our non-relativistic case.
\end{itemize}

Although the parameters $d_{1}$ and $d_{2}$ has not been directly associated
with the electron spin, it should be noted that these two quantities can
also be involved when determining the relative importance of spin for a
specific wave-mode. In some cases it is a combination of the basic
parameters $d_{1}-d_{6}$ that gives the ratio of the spin and classical
contribution. In order to illustrate the situation further, we will briefly
discuss some specific examples.

\textbf{Magnetosonic waves: }For usual ideal MHD waves there are the shear
Alfv\'en wave with $E_{x}\neq 0$, and the fast and slow magnetosonic waves
with $E_{y,z}\neq 0$. The latter wave mode is determined by the components $%
\sigma _{yz}$ ($=-\sigma _{zy}^{\ast }$)\thinspace , $\sigma _{yy}$ and $%
\sigma _{zz}$. Assuming that the plasma is strongly magnetized or has a low
temperature, we consider the case when $d_{3}$ approaches unity, but all
other quantum parameters are small. Considering the MHD-regime it is then
found that the ratio of the spin contribution to $\sigma _{yz}$ (from ($%
Y_{ij}^{(\text{sp})}$)) to the classical contribution (from ($Y_{ij}^{(\text{%
cl})}$)) is $\sigma _{yz}^{(\mathrm{sp})}/\sigma _{yz}^{(\mathrm{cl})}\sim
d_{3}^{2}$. For a low-beta plasma this mainly affects the slow magnetosonic
mode, but for a high beta plasma both the fast and slow magnetosonic mode is
much affected in this regime. This example mainly applies to strongly
magnetized plasmas that occur in astrophysics, e.g.\ pulsars and magnetars.
If experiments on ultra-cold plasmas \cite{Ultracold} were extended to
include strongly magnetized plasmas, this could also be a regime of
relevance.

\textbf{Ion-cyclotron waves:} When ordinary MHD waves (e.g.\ shear Alfv\'en
waves and compressional Alfv\'en waves) have their wavelengths shortened to $%
k_{z}\sim \omega _{ci}/c_{A}$ and $k_{\bot }\sim \omega _{ci}/c_{A}$ where $%
\omega _{ci}$ is the ion-cyclotron frequency, the frequencies are increased
up to $\omega \sim \omega _{ci}$, and the waves become dispersive. For
arbitrary directions of propagation, all components of the conductivity
tensor are of relevance, although the classical expression for $\sigma _{xz}$
(and $\sigma _{zx}$) is usually possible to put to zero. For definiteness we
will make a comparison of $\sigma _{xx}$, but other components would work as
well. We consider a regime of relatively high density, but with a magnetic
field that is not particularly strong. For example, $n_{0}=10^{21} \mathrm{cm%
}^{-3}$ and $B_{0}=10^{-4} \mathrm{T}$. We note that this regime means that $%
c_{A}^{2}\ll c^{2}$ and for a moderate temperature (such that the Bessel
functions can be Taylor expanded to lowest order) the only large quantum
parameter is $d_{5}$ (and possibly $d_{4}$). Evaluating the classical part
of $\sigma _{xx}$ (that comes from the ions) and the spin part of the same
component (that comes from the terms proportional to $X_{ij}^{(\text{sp})}$%
(or $Z_{ij}^{(\text{sp})})$ with $a=\pm 1$ such that the relevant
denominators in (\ref{sigma}) becomes $\omega -k_{z}v_{z}\pm (\omega
_{ce}-\omega _{cg})$, we find the ratio $\sigma _{xx}^{(\mathrm{sp})}/\sigma
_{xx}^{(\mathrm{cl})}=d_{5}^{2}$, where we have used that $\left\vert \omega
_{ce}-\omega _{cg}\right\vert $ is of the order of the ion-cyclotron
frequency $\omega _{ci}$. In general all of the different ion-cyclotron
modes are much affected when $d_{5}\rightarrow 1$ \cite{Estimatate-note}.

\textbf{Whistler waves:} Next we focus on the case of whistler waves, $%
\omega \lesssim \left\vert \omega _{ce}\right\vert $, where the electron
conductivity is dominant also for the classical contribution. Furthermore we
let $k_{\bot }v_{\mathrm{th}}/\left\vert \omega _{ce}\right\vert \ll 1$,
such that the Bessel functions can be expanded, and we note that we are
considering the regime of short perpendicular wavelengths, $k_{\bot
}<\left\vert \omega _{pe}\right\vert /c$. The most relevant components here
are $\sigma _{xx}$, $\sigma _{yy}$ as well as $\sigma _{xy}$. The dominant
spin contribution comes from the $\sigma _{xy}$ component of $Y_{ij}^{(\text{%
sp})}$, which should be compared to the $\sigma _{xy}$ component of $%
Y_{ij}^{(\text{cl})}$. \ Using the thermodynamic equilibrium relation $%
(n_{0+}-n_{0-})/(n_{0+}+n_{0-})=\tanh (\mu _{e}B_{0}/k_{B}T)\approx $ $\mu
_{e}B_{0}/k_{B}T$, we then find the ratio $\sigma _{xy}^{(\mathrm{sp}%
)}/\sigma _{xy}^{(\mathrm{cl})}=d_{3}d_{5}$, and similar comparisons can be
made for other conductivity components in the whistler regime. Note that for
a non-relativistic Alfv\'en velocity we can simplify $d_{3}d_{5}=$ $%
d_{6}(\hbar \omega _{p}/k_{B}T)$. Since it is the \textit{thermodynamic
temperature} that enters here (which determines the relative population of
the two spin states), rather than the Fermi temperature, we can very well be
in the non-relativistic regime $d_{6}\ll 1$ simultaneously as $d_{6}(\hbar
\omega _{p}/k_{B}T)\sim 1$. For example, a plasma with density $n_{0}=10^{30}%
\mathrm{cm}^{3}$ and $T=10^{5}\mathrm{K}$ is nonrelativistic to a good
approximation (i.e.\ $d_{6}\ll 1$) at the same time as $d_{3}d_{5}\sim 1$.

The above brief examples illustrates some possibilities to get conductivity
components due to the spin that are comparable to the classical ones in
magnitude. Note that it is hard to give simple general guidelines when spin
effects are important. For example, for a given temperature we may need a
sufficiently strong magnetic field for spin to be important. We saw this in
the example with magnetosonic waves, where the parameter $d_{3}$ should be
of order unity. On the other hand, for a given density, and for certain wave
modes, we may need a sufficiently \textit{weak} magnetic field for spin to
be important, as we saw in the example with ion-cyclotron waves involving
the parameter $d_{5}$.

Furthermore, sometimes the spin terms are small compared to the classical
terms, but still give a significant contribution. The reason is that a
number of small but qualitatively new effects from the spin can be added to
the dispersion relation. Thus, in order to get a somewhat deeper
understanding of the classical - quantum mechanical transition, than mere
estimates of magnitude can give, we will present a specific calculation in
more detail. In the example below, we follow the assumptions made in Ref.\ 
\cite{BMZ-2009}. We then focus on the regime $k_{\bot }v_{th}/\left\vert
\omega _{ce}\right\vert \ll 1$, such that the Bessel functions can be
expanded and consider $d_{3}\ll 1$. As pointed out previously, the latter
strong inequality condition holds for most plasmas, except for some strongly
magnetized astrophysical objects \cite{Astro}. Furthermore, to be specific,
we consider waves with a polarization $\mathbf{E}=E_{y}\widehat{\mathbf{y}}$
and $\mathbf{B}=B_{x}\widehat{\mathbf{x}}+B_{z}\widehat{\mathbf{z}}=$ $%
(k_{z}E_{y}/\omega )\widehat{\mathbf{x}}-(k_{\bot }E_{y}/\omega )\widehat{%
\mathbf{z}}$. \ For this polarization to be possible we must have $\sigma
_{xy},$ $\sigma _{xz}\ll \sigma _{yy}$, which can be verified \textit{a
posteriori}. \ We keep the standard classical terms (see e.g.\ Ref.\ \cite%
{Swanson}) up to order $\omega _{ce}^{-2}$ in an $1/\omega _{ce}$-expansion
(for both ions and electrons), where the dimensionless expansion parameters
are considered to be $\omega /\omega _{ce}$, $k_{z}v_{th}/\omega _{ce}$ and $%
k_{\bot }v_{th}/\omega _{ce}$. The spin effects are
assumed to be smaller, due to $\hbar \left\vert \omega _{ce}\right\vert
/k_{B}T\ll 1$, and we also consider the case $\omega /\left\vert \omega
_{ce}-\omega _{cg}\right\vert _{c}\ll 1$. Accordingly we only include spin
terms to zero:th order in the $1/\omega _{ce}$-expansion. With these
prerequisites the only spin effect that survives is the $z$-component of the
magnetic dipole force, and its corresponding modification of the $y-$%
component of the magnetization current. For the given polarization the
dispersion relation then reads 
\begin{equation}
\omega ^{2}-k^{2}c^{2}-\frac{i\omega }{\varepsilon _{0}}\sigma _{yy}=0
\label{dr-first}
\end{equation}%
where $\sigma _{yy}$ is given by 
\begin{equation}
\sigma _{yy}=i\varepsilon _{0}\omega \frac{\omega _{pi}^{2}}{\omega _{ci}^{2}%
}+i\varepsilon _{0}\frac{\hbar ^{2}k_{z}^{2}k_{\bot }^{2}\omega _{pe}^{2}}{%
4\omega m_{e}^{2}}\int \frac{\widehat{F}_{0}}{\left( \omega
-k_{z}v_{z}\right) ^{2}}dv_{z}  \label{Eq. New-DR}
\end{equation}%
where all integrations except that over $dv_{z}$ has been carried out, and
the normalization of the rescaled distribution function $\widehat{F}_{0}$ is
given by $\int \widehat{F}_{0}dv_{z}=1$. In Eq.\ (\ref{Eq. New-DR}) the
first term of the right hand side is the classical contribution from the ion
free current, whereas the second term is due to the electron magnetization
current. The dispersion relation (\ref{dr-first}) now immediately reduces to 
\begin{equation}
\omega ^{2}=\frac{k^{2}c^{2}}{1+\omega _{pi}^{2}/\omega _{ci}^{2}}\left[
1-\sin ^{2}\alpha \cos ^{2}\alpha \frac{\hbar ^{2}k^{2}\omega _{pe}^{2}}{%
4m_{e}^{2}c^{2}}\int \frac{\widehat{F}_{0}}{\left( \omega -k_{z}v_{z}\right)
^{2}}dv_{z}\right]   \label{Eq-DR-main}
\end{equation}%
where $\sin ^{2}\alpha =k_{\bot }^{2}/k^{2}$. Firstly, we note that for
ordinary densities in space and laboratory plasmas, the influence of the
real value of $\omega $ is small and thus we can omit the spin term as well
as the first term of Eq.\ (\ref{dr-first}) and write the standard dispersion
relation for compressional Alfv\'en waves $\omega ^{2}=k^{2}c_{A}^{2}$, where
we have assumed $c_{A}\ll c$ (or $\omega _{pi}^{2}/\omega _{ci}^{2}\gg 1$).
Furthermore, for $\mathrm{Re}(\omega )$ to be significantly changed by the
spin effects, we need high densities and/or low temperatures such that $%
\hbar ^{2}\omega _{pe}^{2}/m_{e}^{2}c^{2}v_{t}^{2}$ approaches unity.
However, increasing the density to fulfill this, the minimum velocity spread
will also increase, and when this parameter approaches unity we have also
reached the regime of a relativistic Fermi velocity. Nevertheless, one can
see that the spin has a certain significance even if the dimensionless
parameter of this example is much smaller than unity. Firstly, the spin term
makes the dispersion relation slightly anisotropic, since the group velocity
is not parallel to the wave vector anymore. Secondly, in the given
approximation the group dispersion comes solely from the spin effect.
Finally, provided the omitted cyclotron resonances lies far out in the
thermal tail, the imaginary contribution is determined by the spin term, and
is given by 
\begin{equation}
\gamma \equiv \mathrm{Im}(\omega )=\left( \frac{k^{2}c^{2}}{1+\omega
_{pi}^{2}/\omega _{ci}^{2}}\right) ^{1/2}\frac{\pi \sin ^{2}\alpha \hbar
^{2}\omega _{pe}^{2}}{2m_{e}^{2}c^{2}v_{te}^{2}}\exp \left( -\frac{(\mathrm{%
Re}\omega )^{2}}{k_{z}^{2}v_{t}^{2}}\right) 
\end{equation}%
when the influence of the spin term is small. Thus we can conclude that
there are several possibilities for spin to be of significance, even if all
dimensionless parameters determining the magnitude of the spin contribution
are much smaller than unity.

As a final comment, let us make a few comments regarding the terms in Eq.\ (%
\ref{sigma}) with denominators $\omega -k_{z}v_{s}\pm (\omega _{ce}\pm
\omega _{cg})$. As pointed out in Ref.\ \cite{Brodin2008}, these terms may
give raise to wave modes with frequencies $\omega \approx \left\vert \omega
_{cg}-\omega _{ce}\right\vert $, which is a completely new effect due to
spin,  that can survive even when all dimensionless parameters $d_{1}-d_{6}$
are small. In the next section we will demonstrate from the general result,
Eq. (\ref{sigma}), that the presence of the denominators $\omega
-k_{z}v_{s}\pm (\omega _{ce}\pm \omega _{cg})$ has further consequences for
the transition from wave damping to instability. 


\section{Spin instability - an example}

In order to illustrate the usefulness of (\ref{sigma}), we will evaluate the
imaginary contribution associated with the denominators $\omega
-k_{z}v_{z}\pm (\omega _{cg}-\omega _{c})$. To compute such terms
explicitly, we must pick a specific form of the unperturbed distribution
function. As we here have not been interested in classical instabilities,
that can be induced by a velocity-space non-equilibrium, we have assumed the
distribution function to be isotropic, i.e. $f_{0}=f_{0}(v^{2})$. However,
as we will see below, this does not remove all possibilities for free energy
to be present in the background distribution function. Specifically, in case
the number of particles in the two spin states does not correspond to
thermodynamic equilibrium, we may find that certain wave modes may become
unstable. To illustrate this idea we let the unperturbed distribution
function $F_{0\pm }$ be of the form 
\begin{equation*}
F_{0\pm }=n_{0}\left( \frac{m_{e}}{2\pi k_{B}T_{\text{kin}}}\right)
^{3/2}\exp \left[ -\frac{m_{e}v^{2}}{2k_{B}T_{\text{kin}}}\right] G_{\pm
}(T_{\text{sp}})
\end{equation*}%
where $G_{\pm }(T_{\text{sp}})$ is proportional to the number of particles
in spin-up and spin-down states respectively and can be written as 
\begin{equation*}
G_{\nu }(T_{\text{sp}})\equiv \frac{\exp (\nu \mu _{e}B_{0}/k_{B}T_{\text{sp}%
})}{\exp (\mu _{e}B_{0}/k_{B}T_{\text{sp}})+\exp (-\mu _{e}B_{0}/k_{B}T_{%
\text{sp}})}
\end{equation*}%
where we have introduced the parameter $T_{\mathrm{sp}}$ that can be
interpreted as a spin temperature, in addition to the kinetic temperature $%
T_{\mathrm{kin}}$ of the ordinary Maxwellian velocity dependence. If $T_{%
\mathrm{sp}}=T_{\mathrm{kin}}=T$ we get a thermodynamic equilibrium
distribution with the common temperature $T$, and if we let $T_{\mathrm{sp}%
}\rightarrow \infty $, we get the same number of particles in the two spin
states. We note that for most plasmas, the energy difference between the
high-energy and low-energy spin state is small ( i.e. typically $\mu
_{e}B_{0}/k_{B}T\ll 1$), but nevertheless even a small energy energy
difference can be crucial, as we will see below.


Next we assume that the plasma parameters ($B_{0},n_{0},T_{\mathrm{kin}}$)
correspond to a classical regime, such that the real part of the frequency $%
\omega _{r}(k_{\bot },k_{z})$ is given by the classical terms in (\ref{sigma}%
) to a good approximation. Furthermore, we assume that all resonant electron
velocities $v_{z}=\pm (\omega -n\omega _{c})/k_{z}$ corresponding to the
classical terms lies very far out in the thermal tail. The dominant
imaginary contribution to the dispersion relation may then come from the
spin terms $X_{ij}^{(sp)}$ and $Z_{ij}^{(sp)}$ with resonant particle
velocities $v_{z}=(\omega \pm \left\vert \Delta \omega _{c}\right\vert
)/k_{z}$ (corresponding to $a=\pm 1$ in the conductivity tensor (\ref{sigma}%
)) with $\left\vert \Delta \omega _{c}\right\vert =\left\vert \omega
_{cg}-\omega _{ce}\right\vert $. 
Furthermore, letting the Larmor radius be smaller than $k_{\bot }^{-1}$
(such that the Bessel functions can be Taylor expanded), we can compute the
pole contributions from $X_{ij}^{(sp)}$ and $Z_{ij}^{(sp)}$ with the help of
well known properties of the plasma dispersion function; ${\rm Im}Z(\zeta
)=i\pi ^{1/2}\exp (-\zeta ^{2})$. Since the lower value of the resonant
velocity occurs for $v_{z}=(\omega -\left\vert \Delta \omega _{c}\right\vert
)/k_{z}$  we concentrate on the contribution from $X_{ij}^{(sp)}$ (the
contribution from $Z_{ij}^{(sp)}$ at $v_{z}=(\omega +\left\vert \Delta
\omega _{c}\right\vert )/k_{z}$ can be computed analogously). The result is
then 
\begin{eqnarray}
\text{Im}(\sigma _{ij}) &=&n_{0}\sqrt{\frac{2\pi k_{B}T_{\text{kin}}}{m_{e}}}%
\frac{\mu _{e}^{2}k_{\bot }^{2}}{4\hbar k_{z}\omega _{c}^{2}\omega }\left(
-\tanh \left( \frac{\mu _{e}B_{0}}{k_{B}T_{\text{sp}}}\right) +\frac{\hbar
\omega _{ce}}{2k_{B}T_{\text{kin}}}\right) \exp \left[ \frac{-m_{e}\left(
\omega -\left\vert \Delta \omega _{c}\right\vert \right) ^{2}}{%
2k_{z}^{2}k_{B}T_{\text{kin}}}\right] M_{ij}  \notag \\
&=&A_{ij}(T_{\text{kin}},n_{0},B_{0})\left( \tanh \left( \frac{g}{2}\tilde{a}%
\tilde{T}\right) -\tilde{a}\right)   \label{Im-sigma}
\end{eqnarray}%
where we have noted that $\sum_{\nu }G_{\nu }=1$ and $\sum_{\nu }\nu G_{\nu
}=\tanh (\mu_{e}B_{0}/k_{B}T_{\mathrm{sp}})$, and we have defined $\tilde{a}%
=\frac{\hbar \left\vert \omega _{ce}\right\vert }{2k_{B}T_{\mathrm{kin}}}$
and $\tilde{T}=T_{\mathrm{kin}}/T_{\mathrm{sp}}$. In Fig.\ 1 we plot $\text{Im}(\sigma _{ij})$ as a function of $\tilde{T}$ for some different values of $\tilde{a}$. 
\begin{figure}\label{fig:instability}
\includegraphics[width=0.5\textwidth]{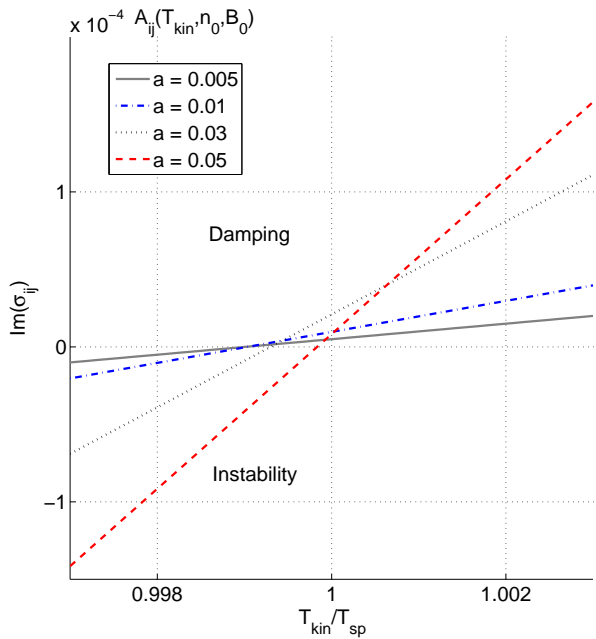}
\caption{(Color online) The imaginary contribution of $X_{ij}^{(sp)}$ to $\sigma_{ij}$ plotted as a function of $T_{\mathrm{sp}}/T_{\mathrm{kin}}$ for different values of $\tilde a$. A positive value corresponds to damping while a negative value gives rise to a instability.}
\end{figure}
For the case of thermodynamic equilibrium, $T_{\mathrm{sp}}=T_{\mathrm{kin}}$, the fact that $
g/2 >1$ ensures that $\text{Im}(\sigma _{ij})>0$ (at least for not too strong
magnetic fields, i.e.\ small values of $\tilde{a}$). In this case the
contribution to the conductivity tensor from the poles gives raise to a
small damping rate. This applies independently of the polarization of the
mode under consideration. However, we need only perturb the spin temperature
a little from equilibrium ($T_{\mathrm{sp}}>T_{\mathrm{kin}}$) for the
system to become unstable. The weak damping then turns to a weak growth rate
of all modes where the resonant velocities $v_{z}=(\omega -\Delta \omega
_{c})/k_{z}$ does not fall too far out in the tail of the distribution. We
stress that this may apply to all sorts of ion-cyclotron modes, as $\Delta
\omega _{c}$ is of the same order as the ion-cyclotron frequency. It is not
hard to imagine situations where the plasma state deviates from
thermodynamic equilibrium, in which case the sign of the pole contribution
may change. For example, if particles are flowing from a region of weak
magnetic field to a region of strong magnetic field, the inflowing particles
will be more evenly distributed between spin-up and spin-down states than
that of the local thermodynamic state. Hence the spin-distribution will be
more evenly distributed than for $T_{\mathrm{sp}}=T_{\mathrm{kin}}$, i.e.\
we will get $T_{\mathrm{sp}}>T_{\mathrm{kin}}$. The same situation will also
occur if the external magnetic field is increased slowly, but at a rate
slightly faster than the spin-states can relax to thermodynamic equilibrium.

It may be noted that if $\tilde{a}$ is sufficiently large (corresponding to
a strongly magnetized systems), the result (\ref{Im-sigma}) predicts that
the system will be unstable already at thermal equilibrium. However, as 
$\tilde{a}$ increases, the effect of Landau quantization will become
important (which affects the perpendicular kinetic energy distribution, see
e.g. \cite{Jens2009}), and hence the background distribution function will
no longer be isotropic. The validity of Eq. (\ref{Im-sigma}) is therefore
restricted to small values of $\tilde{a}$. Finally, let us make an estimate
for the growth rate for the case of ion-cyclotron Alfv\'en waves with $\omega
\sim \omega _{ci}\sim k_{z}c_{A}\sim k_{\bot }c_{A}$. We let the the
temperature ratio be well above the threshold for instability (i.e. $T_{%
\mathrm{sp}}/T_{\mathrm{kin}}\sim 2$), and let $\tilde{a}\ll 1$ (such that
the isotropic distribution function is justified). The normalized growth
rate $\gamma /\omega $ found from \ref{Im-sigma} is then of the order  
\begin{equation}
\frac{\gamma }{\omega }\sim \frac{\hbar ^{2}\omega _{pe}^{2}}{%
m_{e}m_{i}c_{A}^{2}}  \label{growth-rate}
\end{equation}%
if the let the resonance approach the bulk of the distribution in order to
find the maximum groth rate.


\section{Conclusion}

In the present paper we have studied the regime of long spatial scales in an
otherwise fully quantum mechanical kinetic model. The assumption of long
spatial scales (compared to the characteristic de Broglie wavelength) makes
the model reminiscent of semi-classical theory. In particular, the
distribution function behaves as if particles are effectively localized
spatially, and unlike the Wigner function the distribution function is
always positive in our case. Still, the impossibility for the spin vector of
individual particles to have a localized probability distribution in spin
space keeps certain quantum features in the evolution equation (\ref{Vlasov}%
). The linearized theory is solved in a magnetized plasma for a homogeneous
background, which leads to the general conductivity tensor, Eq.\ (\ref{sigma}%
), that includes both the contribution from the free current density as well
as that from the magnetization current due to the spin. This is the main
result of the paper. The main restriction in applying Eq.\ (\ref{sigma}) is
that the velocity distribution is assumed to be isotropic. Specifically we
have used $f_{0}=f_{0}(v^{2},\theta _{s})=(1/4\pi)\left[F_{0+}(v^{2})(1+\cos
\theta _{s})\right.$$+\left.F_{0-}(v^{2})(1-\cos \theta _{s})\right]$. Note
that, by contrast, the spin distribution of the unperturbed distribution
function is the most general time-independent solution for a constant
magnetic field background.

In section III we have discussed the quantum-classical boundary, which is
considerably more complicated when spin effects are included. In particular,
as deduced from the dimensionless parameter $d_{3}-d_{6}$, spin effects may
in certain cases remain also in a plasma of modest temperature and density.
Furthermore, effects due to the new types of wave-particle resonances may be
of signficance independent of the dimensioneless parameters. In section IV,
we have found that such resonances is one of the key ingredients in a new
type of \ instability, somewhat reminiscent of the Weibel-instability. As is
wellknown (see e.g. \cite{Swanson}), in the case of Weibel-instabilites  the
instabilities arise from a difference in perpendicular and parallell kinetic
temperatures. Here, the deviation in the spin temperature from a common
(isotropic) kinetic temperature is the source of the instability.
Furthermore we note that a very small deviation of $T_{\mathrm{kin}}/T_{%
\mathrm{sp}}$ from unity is sufficient to drive an instability. 

Although the present result (\ref{sigma}) is relatively general, there is
still several interesting extensions that could be made;

\begin{enumerate}
\item Generalization to an arbitrary (non-isotropic) background velocity
distribution.

\item Inclusion of the spin-orbit coupling.

\item Using the full evolution of Ref.\ \cite{Jens2009}, including the
short-scale physics (i.e.\ shorter than the characteristic de Broglie
wavelength).

\item Inclusion of collisional effects.
\end{enumerate}

Furthermore, a thorough evaluation of (\ref{sigma}), that must be done
numerically, could possibly reveal new and interesting possibilities,
involving e.g.\ new wave modes.


\appendix

\section{Some useful relations}

\label{appendix:relations} The integrals (\ref{inthelp1})-(\ref{inthelp3})
are very useful when performing the $\varphi _{s}$ and $\varphi _{v}$
integrations in Eqs.\ (\ref{Iab}), (\ref{Jf}) and (\ref{Jm}), 
\begin{subequations}
\begin{equation}  \label{inthelp1}
\int_{0}^{2\pi }d\varphi \exp (in\varphi )\cos (\varphi )=\left\{ 
\begin{array}{ccc}
\pi & \mathrm{for} & n=\pm 1 \\ 
0 & \mathrm{for} & n\neq \pm 1,%
\end{array}%
\right.
\end{equation}%
\begin{equation}  \label{inthelp2}
\int_{0}^{2\pi }d\varphi \exp (in\varphi )\sin (\varphi )=\left\{ 
\begin{array}{ccc}
\pm i\pi & \mathrm{for} & n=\pm 1 \\ 
0 & \mathrm{for} & n\neq \pm 1,%
\end{array}%
\right.
\end{equation}%
\begin{equation}  \label{inthelp3}
\int_{0}^{2\pi }d\varphi \exp (in\varphi )=\left\{ 
\begin{array}{ccc}
2\pi & \mathrm{for} & n=0 \\ 
0 & \mathrm{for} & n\neq 0.%
\end{array}%
\right.
\end{equation}
When performing the $\theta _{s}$ integration in going from Eqs.\ (\ref{Jf})
and (\ref{Jm}) to (\ref{sigma}), it is useful to note that integrals on the
form 
\end{subequations}
\begin{equation}  \label{inthelp4}
\int_{0}^{\pi }d\theta \left(\cos \theta\right)^{n} \left(\sin
\theta\right)^{m}
\end{equation}%
vanish for all odd integer values of $n$ independent of the integer value of 
$m$.


\section{The Hermitian structure of the conductivity tensor}

\label{App-Hermitian} The Hermitian structure of the conductivity tensor ($%
\sigma _{ij}=-\sigma _{ij}^{\ast }$) is not transparent from Eqs.\ (\ref{Jf}%
) and (\ref{Jm}). The conductivity tensor will for instance not be
recognized to have a Hermitian structure until after the $\theta _{s}$%
-integration is performed. Below we list some further identities that have
been used in obtaining $\sigma _{ij}$ on the form written in Eq.\ (\ref%
{sigma}).

To match the terms $\sigma_{xy}$ with $\sigma_{yx}$, it is found that an
integral of the form 
\begin{eqnarray}
\sum_{a=-\infty}^\infty \int_0^{\infty} \frac{a\mathcal{J}_a^2}{\omega -
k_zv_z-a\omega_{ce}}\frac{\partial f_0}{\partial v^2}v_\bot dv_\bot
\end{eqnarray}
must match the integral 
\begin{eqnarray}
&& \sum_{a=-\infty}^\infty \int_0^{\infty} \frac{a}{\omega}\frac{%
(a\omega_{ce} + k_zv_z)\mathcal{J}_a^2}{\omega - k_zv_z-a\omega_{ce}}\frac{%
\partial f_0}{\partial v^2}v_\bot dv_\bot  \notag \\
&=& \sum_{a=-\infty}^\infty \int_0^{\infty} \left( -a\mathcal{J}_a^2 + \frac{%
a\mathcal{J}_a^2}{\omega - k_zv_z-a\omega_{ce}}\right)\frac{\partial f_0}{%
\partial v^2}v_\bot dv_\bot.
\end{eqnarray}
For the terms to match, the following integral must vanish; 
\begin{eqnarray}  \label{Herm-int}
\sum_{a=-\infty}^\infty \int_0^{\infty} a\mathcal{J}_a^2\frac{\partial f_0}{%
\partial v^2}v_\bot dv_\bot &=& \frac{1}{2}\sum_{a=-\infty}^\infty
\int_0^{\infty} a\mathcal{J}_a^2\frac{\partial f_0}{\partial v_\bot} dv_\bot
\notag \\
&=& -\frac{1}{2} \int_0^{\infty}\sum_{a=-\infty}^\infty a\mathcal{J}_a(%
\mathcal{J}_{a-1} - \mathcal{J}_{a+1})f_0 dv_\bot  \notag \\
&=& -\frac{1}{2} \int_0^{\infty}\sum_{n=1}^\infty \left[n\mathcal{J}_n (%
\mathcal{J}_{n-1} - \mathcal{J}_{n+1}) - n\mathcal{J}_{-n} (\mathcal{J}%
_{-(n+1)} - \mathcal{J}_{-(n-1)})\right] f_0 dv_\bot  \notag \\
&=& 0
\end{eqnarray}
where we have integrated by parts, summed over all integers $a$ and used the
relation $\mathcal{J}_{-n}=(-1)^n\mathcal{J}_n$. Thus, since the integral (%
\ref{Herm-int}) indeed vanish, we have $\sigma_{xy}=-\sigma_{yx}^*$ as
expected.

In a similar manner, we find that the following integral must vanish for 
$\sigma_{yz}$ to match $\sigma_{zy}$; 
\begin{eqnarray}
\int_{-\infty}^{\infty} v_z\frac{\partial f_0}{\partial v^2}dv_z =0.
\end{eqnarray}
This is trivially satisfied under the assumption $f_0(v^2)$ since this is a
product of an odd and an even function integrated from $-\infty$ to $\infty$.


\section{The short Larmor radius limit}

\label{sec:SLRL-limit} Below we consider the short Larmor radius limit of
the conductivity tensor (\ref{sigma}), i.e.\ the case where the Bessel
argument is small $k_\bot v_\bot/\left|\omega_{ce}\right| \ll 1$. Since we have assumed
an isotropic distribution for the plasma ($f_0(v^2)$), we note that 
\begin{eqnarray}
\frac{\partial F_0}{\partial v^2} = \frac{1}{2v_z}\frac{\partial F_0}{%
\partial v_z}.  \notag
\end{eqnarray}
With this noted, we can integrate by parts with respect to $v_z$ to
eliminate the derivatives of $F_0$ in the conductivity tensor (\ref{sigma}).
We encounter integrals on the form 
\begin{eqnarray}
\int_{-\infty}^\infty \frac{v_z^n}{\Omega - k_zv_z}\frac{\partial F_0}{%
\partial v^2}dv_z &=& \frac{1}{2}\int_{-\infty}^\infty \frac{v_z^{n-1}}{%
\Omega - k_zv_z}\frac{\partial F_0}{\partial v_z}dv_z  \notag \\
&=& - \frac{1}{2}\int_{-\infty}^\infty F_0 \left(\frac{(n-1)v_z^{n-2}}{%
\Omega - k_zv_z} + \frac{k_zv_z^{n-1}}{(\Omega - k_zv_z)^2}\right)dv_z
\end{eqnarray}
where $n$ may take the integer values $n=0,1,2$. Here $\Omega \equiv \omega
- a\omega_{ce} - b\omega_{cg}$ with the integer values of $a$ and $b$ not
being specified. Next, we expand the Bessel functions in $v_\bot
k_\bot/\omega_{ce}$ and perform the summation over all integer values of $a$%
. Terms proportional to $k_\bot v_\bot/\omega_{ce}$ are considered small and
are dropped. The conductivity tensor then becomes: 
\begin{eqnarray}  \label{sigma-SLRL}
\sigma_{ij} = \sum_{\nu=+,-} \left[x_{(\nu)ij}^{(\text{sp})} + y_{(\nu)ij}^{(%
\text{cl})} + y_{(\nu)ij}^{(\text{sp})} + z_{(\nu)ij}^{(\text{sp})}\right]
\end{eqnarray}
where 
\begin{equation}
y_{(\nu)ij}^{(\text{cl})}= \frac{q_e^2}{m_e}\int F_{0\nu}\left[ \tfrac{\omega}{%
(\omega - k_zv_z)^2}\times\left( 
\begin{array}{ccc}
0 & 0 & 0 \\ 
&  &  \\ 
0 & 0 & 0 \\ 
&  &  \\ 
0 & 0 & i%
\end{array}
\right) + \tfrac{1}{4}\sum_{a=\pm 1} \tfrac{\omega - 2k_zv_z -a\omega_{ce}}{%
(\omega - k_zv_z - a\omega_{ce})^2} \tfrac{ v_\bot^2}{ v_z^2}\times\left( 
\begin{array}{ccc}
-i & a & 0 \\ 
&  &  \\ 
-a & - i & 0 \\ 
&  &  \\ 
0 & 0 & 0%
\end{array}
\right) \right]d^3v  \notag
\end{equation}
is the classical contribution, and the spin contributions are 
\begin{equation}
y_{(\nu)ij}^{(\text{sp})}= \mu_e\frac{q_e}{m_e}\int F_{0\nu} \left[\tfrac{1}{%
(\omega - k_zv_z)^2}\times\left( 
\begin{array}{ccc}
0 & 0 & 0 \\ 
&  &  \\ 
0 & \left(i\frac{\mu_e}{q_e}\frac{k_\bot^2 k_z^2}{\omega} \nu i\frac{1}{2} 
\frac{v_\bot^2}{v_z^2}\frac{k_\bot^2(\omega - 2k_zv_z)}{\omega_{ce}}\right)
& \nu k_\bot k_z \\ 
&  &  \\ 
0 & -\nu k_\bot k_z & 0%
\end{array}
\right)\right.  \notag
\end{equation}

\begin{equation}
\qquad - \tfrac{1}{4}\tfrac{k_\bot^2}{\omega_{ce}}\sum_{a=\pm 1}\tfrac{%
\omega-2k_zv_z-a\omega_{ce}}{(\omega-k_zv_z - a\omega_{ce})^2} \tfrac{%
v_\bot^2}{v_z^2}\times\left.\left( 
\begin{array}{ccc}
0 & -\nu a & 0 \\ 
&  &  \\ 
\nu a & \left(ia \frac{\mu_e}{q_e}\frac{k_\bot^2}{\omega} + \nu i \frac{%
\left(\omega+a\omega_{ce}\right)}{\omega} \right) & 0 \\ 
&  &  \\ 
0 & 0 & 0%
\end{array}
\right)\right]d^3v  \notag
\end{equation}
together with 
\begin{eqnarray}
x_{(\nu)ij}^{(\text{sp})} + z_{(\nu)ij}^{(\text{sp})} &=&
\sum_{b=\pm1}\left\{i\frac{\mu_e^2}{\hbar\omega}\int F_{0\nu} \left[\frac{%
-\nu b}{\omega - k_zv_z - b\omega_{cg}} - \frac{1}{2}\frac{\hbar}{m}\frac{%
k_z^2}{(\omega - k_zv_z - b\omega_{cg})^2} \right.\right.  \notag \\
&+& \left.\left.\sum_{a=\pm 1} a\frac{1}{8}\frac{\hbar}{m_e}\frac{k_\bot^2}{%
\omega_{ce}}\frac{v_\bot^2}{v_z^2}\frac{\omega -
2k_zv_z-a\omega_{ce}-b\omega_{cg}}{(\omega - k_zv_z - a\omega_{ce} -
b\omega_{cg})^2} \right]M^{(b)}_{ij}d^3v\right\}  \notag
\end{eqnarray}
where $M^{(-1)}_{ij}\equiv M_{ij}$ with $M_{ij}$ being defined as before, and $%
M^{(1)}_{ij}\equiv M^*_{ij}$.

We may note that we only encounter four kinds of unique integrals in the
conductivity tensor (\ref{sigma-SLRL}); 
\begin{eqnarray}
&&\int F_{0}\frac{1}{\Omega -k_{z}v_{z}}d^{3}v,  \notag \\
&&\int F_{0}\frac{v_{\bot }^{2}}{v_{z}^{2}}\frac{1}{(\Omega -k_{z}v_{z})^{2}}%
d^{3}v,  \notag \\
&&\int F_{0}\frac{v_{\bot }^{2}}{v_{z}}\frac{1}{(\Omega -k_{z}v_{z})^{2}}%
d^{3}v,  \notag \\
&&\int F_{0}\frac{1}{(\Omega -k_{z}v_{z})^{2}}d^{3}v.  \notag
\end{eqnarray}%
For the special case when the cyclotron resonances are far out in the
thermal tail, the conductivity tensor (\ref{sigma-SLRL}) may be
significantly simplified. This is, however, trivial to obtain from (\ref%
{sigma-SLRL}) and will therefore not be pursued further here.


\begin{thebibliography}{99}
\bibitem{Manfredi-review} G. Manfredi, Fields Inst. Comm. \textbf{46}, 263
(2005).

\bibitem{Shukla-Eliasson-review} P. K. Shukla and B. Eliasson, Phys.-Usp.%
\textbf{\ 53}, 51 (2010).

\bibitem{Haas-2000} F. Haas, G. Manfredi, and M. R. Feix, Phys. Rev. E 
\textbf{62}, 2763 (2000).

\bibitem{Garcia-2005} L. G. Garcia, F. Haas, L. P. L. de Oliviera, and J.
Goedert, Phys. Plasmas \textbf{12}, 012302 (2005).

\bibitem{Shukla-Eliasson-2006} P. K. Shukla and B. Eliasson, Phys. Rev.
Lett. \textbf{96}, 245001 (2006).

\bibitem{Jens2009} J. Zamanian, M. Marklund, and G. Brodin, New J. Phys. 
\textbf{12}, 043019 (2010).

\bibitem{Marklund2007} M. Marklund and G. Brodin, Phys. Rev. Lett. \textbf{98%
}, 025001 (2007).

\bibitem{Classical-quant} G. Brodin, M. Marklund, and G. Manfredi, Phys.
Rev. Lett. \textbf{100}, 175001 (2008).

\bibitem{Marklund-plasmonic} M. Marklund, G Brodin, L. Stenflo and C. S.
Liu, Europhys. Lett. \textbf{84}, 17006 (2008).

\bibitem{Brodin2008} G. Brodin, M. Marklund, J. Zamanian, A. Ericsson, and
P. L. Mana, Phys. Rev. Lett. \textbf{101}, 245002 (2008).

\bibitem{BMZ-2009} G. Brodin, M. Marklund and J. Zamanian p. 280-290 in 
\textit{New developments in nonlinear plasma physics}. Eds. B. Eliasson and
P. K. Shukla, AIP conf. proce. No 1188, (AIP, New York, 2009).

\bibitem{Manfredi-quantum-well} G. Manfredi and P.-A. Hervieux, Appl. Phys.
Lett. \textbf{91}, 061108 (2007).

\bibitem{Atwater-Plasmonics} H. A. Atwater, Sci. Am.\textbf{\ 296}, 56
(2007).

\bibitem{Spintronics} S. A. Wolf, D. Awschalom, R. A. Buhrman, \textit{et al.}, Science \textbf{294}, 1488
(2001).

\bibitem{Astro} C. Kouveliotou, S. Dieters and T. Strohmayer \textit{et al.}%
, Nature \textbf{393}, 235 (1998); D. M. Palmer, S. Barthelmy, and N.
Gehrels, Nature \textbf{434}, 1107 (2005); A. K. Harding and D. Lai, Rep.
Prog. Phys. \textbf{69}, 2631 (2006).

\bibitem{Ultracold} M. P. Robinson, B. Laburthe Tolra, M. W. Noel, T. F. Gallagher, and P. Pillet, Phys. Rev. Lett. \textbf{85}, 4466 (2000).

\bibitem{Solid-density} S. H. Glenzer \textit{et al.}, Phys. Rev. Lett. 
\textbf{98}, 065002 (2007).

\bibitem{Cowley-1986} S. C. Cowley, R. M. Kulsrud, and E. Valeo, Phys.
Fluids \textbf{29}, 430 (1986).

\bibitem{Kulsrud-1986} R. M. Kulsrud, E. J. Valeo, and S. C. Cowley, Nucl.
Fusion \textbf{26}, 1443 (1986).

\bibitem{Swanson} D. G. Swanson, \textit{Plasma Waves} (Taylor \& Francis,
2003).

\bibitem{note1} It should be noted that we have defined the conductivity
tensor to include all contributions to the current density, i.e.\ the
conductivity tensor includes the free current density (due to the Lorentz
force and the magnetic dipole force) as well as the contributions from the
magnetization current density.

\bibitem{Estimatate-note} When performing this estimate, we have ignored the
spin-contribution from the first term in $X_{ij}^{(\text{sp})}$(and $%
Z_{ij}^{(\text{sp})}$). This is valid e.g.\ if there is equal number of
spin-up and spin-down populations, in which case this contribution vanishes.
\end{thebibliography}
\end{document}